\documentclass[11pt,twoside]{article}

\usepackage{asp2010}
\usepackage{graphicx}
\pdfoutput=1

\resetcounters 

\begin{document}

\resetcounters

\title{Observing extreme-mass-ratio inspirals with eLISA/NGO}
\author{Jonathan R Gair$^1$ \affil{$ö1$ Institute of Astronomy, University of Cambridge, Madingley 
Road, Cambridge, CB3 0HA, UK}}
\author{Edward K Porter$^2$ \affil{$ö2$ APC, Univ. Paris Diderot, CNRS/IN2P3, CEA/Irfu, Obs. de Paris, Sorbonne Paris Cit\'{e}, France}}
 
\begin{abstract} 
The extreme-mass-ratio inspirals (EMRIs) of stellar mass compact objects into massive black holes in the centres of galaxies are an important source of low-frequency gravitational waves for space-based detectors. We discuss the prospects for detecting these sources with the evolved Laser Interferometer Space Antenna (eLISA), recently proposed as an ESA mission candidate under the name NGO. We show that NGO could observe a few tens of EMRIs over its two year mission lifetime at redshifts $z \lesssim 0.5$ and describe how the event rate changes under possible alternative specifications of the eLISA design.
\end{abstract}

\section{Introduction} 
The extreme-mass-ratio inspiral (EMRI) of a stellar mass compact object --- a black hole (BH), neutron star (NS) or white dwarf (WD) --- into a massive black hole (MBH), with mass in the range $10^4$--$10^7M_\odot$, in the centre of a galaxy will generate gravitational waves (GWs) in the mili-Hertz frequency range to which space-based detectors, such as LISA~\citep[e.g.][]{2003AdSpR..32.1233D}, will be sensitive. The extreme-mass-ratio ensures that the inspiral proceeds slowly and therefore, from each EMRI system, we expect to observe several hundreds of thousands of waveform cycles generated while the small object is in the strong field region of the spacetime close to the central black hole~\citep{Finn:2000uv}. This emitted radiation encodes a detailed map of the spacetime structure that can be used to measure the parameters of the system to accuracies of a fraction of a percent~\citep{Barack:2004jy} and to test whether these objects are indeed the Kerr black holes predicted by general relativity~\citep{ryan95}. EMRI observations thus have strong potential applications to astrophysics~\citep{GTVemri}, cosmology~\citep{2008PhRvD..77d3512M} and fundamental physics~\citep[see][for a review]{ASReview}.

For the classic Laser Interferometer Space Antenna (LISA) design, the EMRI event rate was estimated to be from a few to as many as several thousand events over the mission lifetime~\citep{emrirate,gairLISA7}, with the range determined primarily by the very uncertain astrophysics of these systems. The withdrawal of NASA funding for the LISA mission in 2010 prompted the redesign of LISA to evolved LISA (eLISA), also called the New Gravitational Observatory (NGO)~\citep{2012CQGra..29l4016A}. This new design, with shorter arm lengths and only four rather than six inter-spacecraft laser links, does not have the same sensitivity to EMRI events as the classic LISA design. In this article we will discuss the sensitivity of eLISA/NGO to EMRIs and provide estimates of the likely EMRI event rate for this modified detector design, comparing it to classic LISA and to two alternative options for a re-scoped mission. We will use the name eLISA to refer to any possible future re-scoped designs of LISA, and the name NGO to refer specifically to the 4-link, 1Gm armlength version submitted as a mission proposal to ESA. 
In Section~\ref{theory} we describe the elements that go into the calculation, including the detector sensitivity, the waveform model and the model for the intrinsic EMRI rate. In Section~\ref{results} we present event rate estimates for EMRIs in eLISA/NGO and compare these to classic LISA. We also discuss the expected masses and redshifts for these detected events. We finish in Section~\ref{discuss} with a short discussion.

\section{Estimating EMRI event rates}
\label{theory}
\subsection{Detector Sensitivity}
The criterion for detection of an EMRI is that the matched-filtering signal-to-noise ratio (SNR), $\rho$, is sufficiently high. This is given by
\begin{equation}
\rho = 4 \int_0^\infty \frac{\tilde{h}^*(f) \tilde{h}(f)}{S_h(f)} {\rm d} f,
\end{equation}
where $h(t)$ is the waveform strain, a tilde denotes the Fourier transform, a star is complex conjugation and $S_h(f)$ is the one-sided power spectral density of noise in the detector. The large dimensionality of the EMRI parameter space makes data analysis for these sources difficult and in previous works it was usually assumed that a high SNR threshold of $\rho_{\rm thresh} \gtrsim 30$ would be required to be confident of a detection~\citep{emrirate,gairLISA7}. This was based on a historic model of EMRI data analysis that used semi-coherent matched filtering. The extraction of EMRIs with SNR as low as $\rho \sim 15$ has subsequently been demonstrated using Markov Chain Monte Carlo techniques~\citep{mldc3}. However, this was for data sets with an unrealistically low density of sources. We therefore adopt a threshold $\rho_{\rm thresh} = 20$ in this work.

EMRIs are long-lived sources, that gradually accumulate SNR over the several years prior to plunge. We therefore characterise EMRI detectability using an observable lifetime~\citep{gairLISA7}. If $T_{\rm m}$ denotes the mission lifetime, we define $\rho(t)$ as the SNR accumulated from a time $t$ before plunge to plunge when $t<T_{\rm m}$ and define $\rho(t)$ to be the SNR accumulated from times $t$ to $t-T_{\rm m}$ before plunge otherwise. The observable lifetime is $T_{\rm obs} = t_2-t_1$, where $t_2$ are the largest/smallest solutions to $\rho(t)=\rho_{\rm thresh}$. We average this observable lifetime for sources at a given redshift over possible choices for the extrinsic parameters of the system to give $ \bar{T}_{\rm obs}(z)$. If the intrinsic EMRI rate per comoving volume is $r_{\rm int}(z)$, then $r_{\rm int}(z) \bar{T}_{\rm obs}(z)$ is the number of events per comoving volume that would be observed at that redshift. Integrating this over redshift and other source parameters provides the final event rate estimate.

The spectral density $S_h(f)$ encodes the assumptions about the detector. We take the spectral density for NGO from~\cite{2012CQGra..29l4016A} and for LISA from~\cite{Barack:2004jy}. NGO is a four-link design, meaning only one independent Michelson response can be constructed from the detector output, as opposed to the six-link/two-detector design of classic LISA. We consider four and six link versions of each, by assuming one or two available independent detectors with the specified sensitivity. We also consider a version of eLISA with four-links but 2 million km arms (NGO has 1 million km arms) with an appropriately scaled sensitivity.

For the waveform model, $h(t)$, we use two different approaches. We consider circular-equatorial EMRIs with waveforms computed using  solutions of the Teukolsky equation, taking results from~\cite{Finn:2000uv} and as implemented in~\cite{gairLISA7}. We also consider generic EMRIs using the analytic kludge model of~\cite{Barack:2004jy}. The Teukolsky results should be more accurate, but the analytic kludge results allow for eccentricity and inclination of the orbits. We average over extrinsic parameters using a sky-averaged sensitivity curve in the former case and by Monte Carlo averaging in the latter case.

\subsection{Astrophysical EMRI rate}
The intrinsic EMRI rate is the product of the number density of black holes per comoving volume, $N$, and the rate of EMRIs occurring in black holes of that mass, $r_{\rm BH}$. Neither of these quantities are well constrained~\citep{ASReview}, so we follow the approach of~\cite{gairLISA7}. We assume that the number density of black holes is flat in logarithm, ${\rm d}N/{\rm d}\ln M = 0.002$Mpc$^{-3}$, and a power-law scaling of the rate per black hole with central black hole mass, $M$, $r_{\rm BH} = 400$Gyr$^{-1} (M/3\times10^6M_\odot)^{-0.17}$~\citep{HopmanLISA7,2011CQGra..28i4017A}. This is the rate for inspirals of black holes. Even for classic LISA the rate of NS or WD inspirals is $\lesssim 1$ over a mission lifetime and there is very little chance of seeing these with eLISA. This mass scaling predicts that lower mass black holes, with $M\sim10^4M_\odot$, will accumulate a significant fraction of their mass from EMRI events, which is unrealistic. We therefore impose the additional constraint that a black hole can acquire no more than $10\%$ of its mass from EMRI events. Imposition of this constraint does not significantly modify the estimated NGO rates.

\section{eLISA/NGO EMRI rates and properties}
\label{results}
In Table~\ref{RateTable} we show the number of events that would be detected over the mission duration (taken to be 2 and 5 years for NGO and LISA respectively) for each of the configurations. These results were computed using the circular-equatorial EMRI model. Event rates were also computed using the analytic kludge model for eccentric EMRIs and these were found to be in very good agreement. These event rates are for the inspirals of stellar mass black holes, assuming all the inspiraling black holes have mass $m=10M_\odot$.  The quoted numbers for classic LISA are smaller than those published elsewhere~\citep{gairLISA7}, which is a consequence of the imposition of the $10\%$ mass fraction cut-off in the intrinsic EMRI rate and that here we consider only events with redshift $z < 1$, while classic LISA is sensitive to EMRI sources at higher redshifts. We see that the EMRI event rate for NGO is of the order of a few tens of events. 
If the black holes tend to be rapidly spinning, the rate could be as much as a factor of two higher and more events involving heavier black holes will be detected. For an up-scoped eLISA with 6-links the increase in EMRI event rate would be a factor of $\sim1.5$, while an up-scope to double the eLISA armlength would lead to a factor $\sim 2$ increase in event rate. Event rate estimates are not the only consideration when contemplating up-scope options, as these also affect parameter estimation accuracies. This is not a significant concern for EMRIs, since the gradual accumulation of SNR means that parameter estimations are almost as good in a 4-link configuration as a 6-link one, but for other source types parameter estimation can be substantially improved by an up-scope to 6-links~\citep{2012CQGra..29l4016A}. If classic LISA were realised, the EMRI event rate would be enhanced further, by up to a factor of $5$-$10$. We must note, however, that these numbers do not account for uncertainties in the astrophysical EMRI rate, which are at least two orders of magnitude~\citep{ASReview}.

\begin{table}
\begin{center}
\begin{tabular}{c|c|c|c|c|c|c|c|c|c|}
&\multicolumn{9}{c}{Black Hole Spin}\\
&\multicolumn{3}{c|}{$a=0$}&\multicolumn{3}{c|}{$a=0.5$}&\multicolumn{3}{c|}{$a=0.9$}\\\hline
Detector&\multicolumn{3}{c|}{No. Events in}&\multicolumn{3}{c|}{No. Events in}&\multicolumn{3}{c|}{No. Events in}\\
&${\cal M}_1$&${\cal M}_2$&${\cal M}_3$&${\cal M}_1$&${\cal M}_2$&${\cal M}_3$&${\cal M}_1$&${\cal M}_2$&${\cal M}_3$\\\hline
NGO&$<1$&15&$<1$&$<1$&19&1&$<1$&45&15\\\hline
6-link NGO&$2$&35&$<1$&$2$&57&3&$2$&70&35\\\hline
2Gm NGO&$5$&45&$2$&$2$&55&5&$3$&95&45\\\hline
4-link LISA&$10$&190&$10$&$10$&210&30&$10$&220&130\\\hline
6-link LISA&$40$&280&$20$&$30$&290&50&$30$&300&160\\\hline
\end{tabular}
\caption{\label{RateTable}EMRI event rates for each detector configuration. Results are shown assuming all black holes have the same spin, but for three different choices of that assumed spin value. Results are also divided up into different mass categories, ${\cal M}_1 \equiv 10^4M_\odot < M < 10^5M_\odot$, ${\cal M}_2 \equiv 10^5M_\odot < M < 10^6M_\odot$ and ${\cal M}_3 \equiv 10^6M_\odot < M$.}
\end{center}
\end{table}

\begin{figure}[!h]
\begin{center}
\begin{tabular}{c}
\includegraphics[angle=0,width=0.8\textwidth]{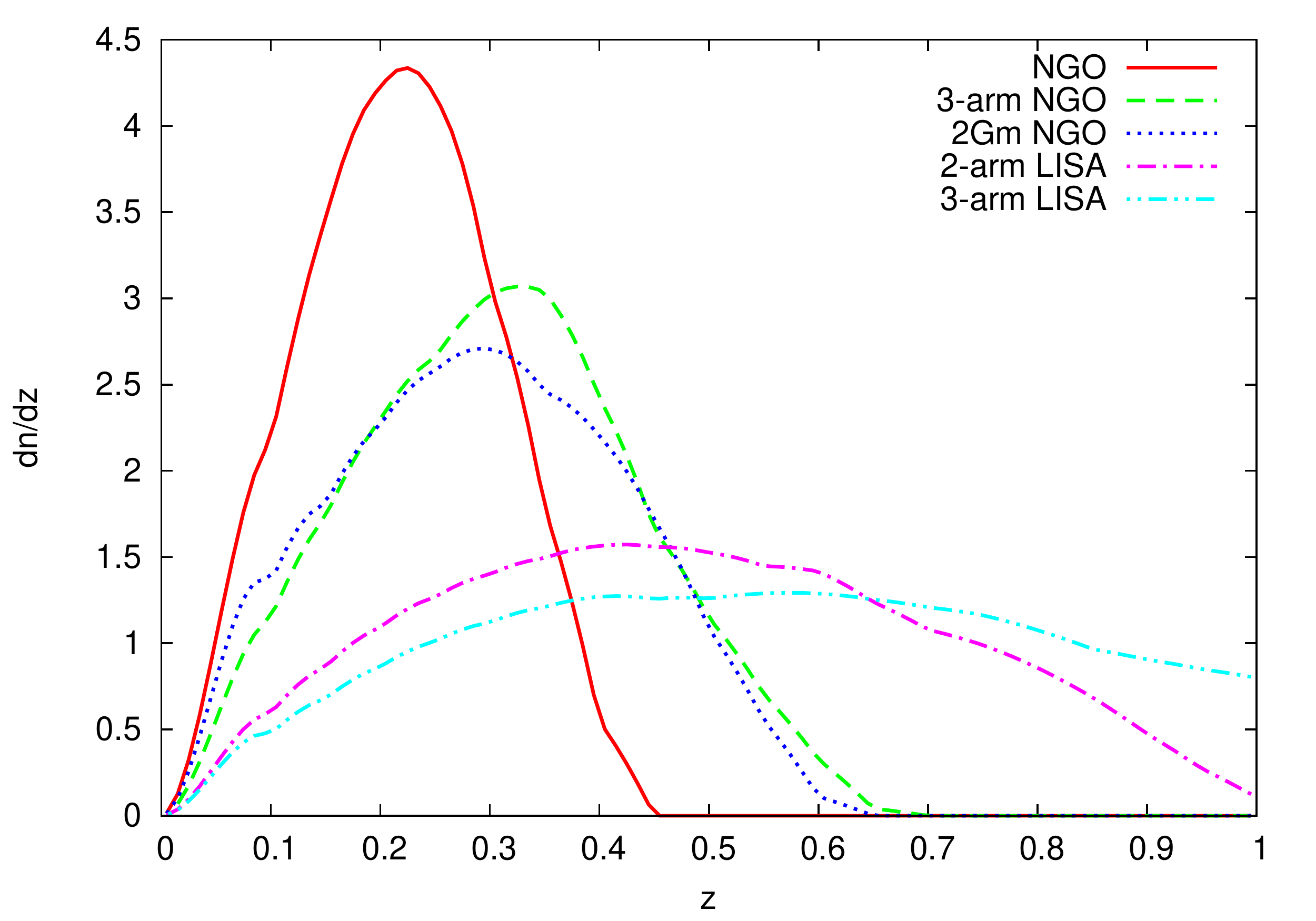}\\
\includegraphics[angle=0,width=0.8\textwidth]{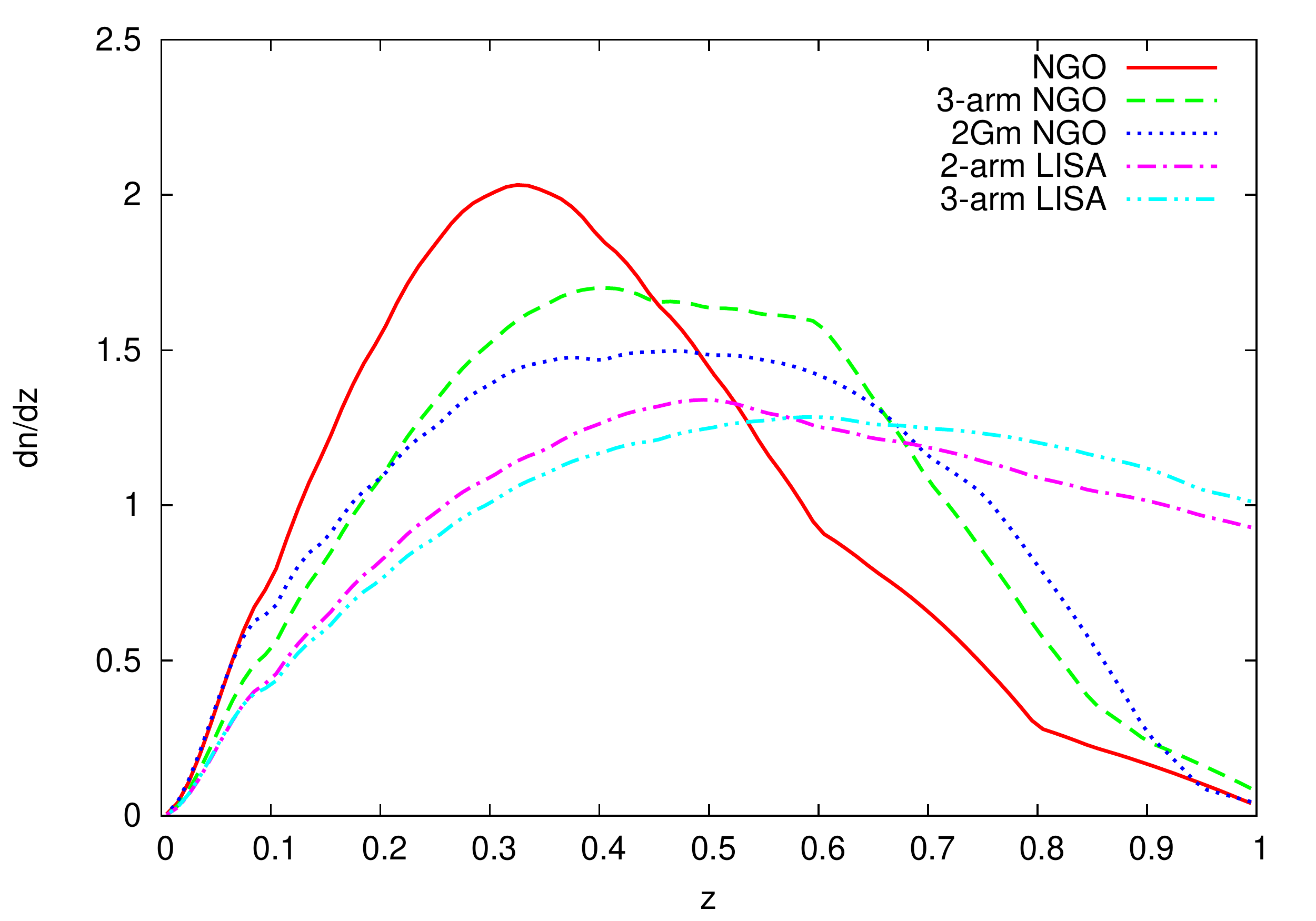}
\end{tabular}
\caption{\label{zdistfig}Redshift distribution of detected EMRI events for each configuration, assuming all MBHs have spin $a=0$ (top panel) or $a=0.9$ (bottom panel).}
\end{center}
\end{figure}

As indicated in the Table, the overwhelming majority of detected EMRI events will have mass between $10^5$ and $10^6$ solar masses and the mass distribution is peaked at $\sim5\times10^5M_\odot$ if all black holes have low spin, or at $\sim7\times10^5M_\odot$ if all black holes have significant spins. The location of this peak does not depend strongly on the detector configuration, although the distribution is somewhat wider for the more sensitive classic LISA configurations. The redshift distribution of detected events is shown in Figure~\ref{zdistfig}. The redshift distribution is fairly broad and peaked at $z \sim 0.2$ for eLISA/NGO, with no events detected above $z\sim0.45$, if black holes are mostly of low spin. If MBHs are more rapidly spinning, the peak is pushed up to $z\sim 0.3$ and the maximum redshift is close to $1$. For up-scoped versions of the detector, the peak and maximum redshift of events are both increased. A 6-link or 2Gm armlength NGO would have a distribution peaked at $z\sim0.3 (0.5)$ for non-spinning (rapidly spinning) MBHs and a maximum redshift $z \sim 0.6 (1)$. Classic LISA would have a distribution peaked at $z\sim0.5 (0.7)$ and maximum redshift $z > 1$. We conclude that the additional events detected in the up-scoped configurations come from the increased redshift range of the detector but the alternative configurations do not open up significant additional regions of the mass parameter space.

\section{Discussion}
\label{discuss}
We have discussed the prospects for detection of EMRI events with eLISA/NGO. The NGO detector could observe a few tens of events over its lifetime, with a typical redshift $z \sim 0.2$ and a typical MBH mass of $M \sim 5\times10^5M_\odot$. Alternative, up-scoped versions of eLISA with a factor of two increase in armlength or six inter-spacecraft laser links would increase the number of events by a factor of two, while an up-scope all the way to classic LISA would provide up to an order of magnitude increase in event rate. These numbers do not include the two or more order of magnitude uncertainty in the astrophysical rates, but the relative event rates in the different configurations should be independent of that uncertainty.

There is extensive literature on the scientific potential of EMRI observations with LISA, so it is natural to ask whether similar objectives can be accomplished with eLISA/NGO. The precision of EMRI parameter estimation comes from accurately tracking the waveform phase over many cycles and therefore every event that is detected will provide precise parameter measurements, irrespective of the exact detector configuration. The impact of the detector design on EMRI science is therefore primarily through the change in event rate. Classic LISA would constrain the local slope of the black hole mass function in the relevant range to $\sim \pm 0.3$, better than current constraints, with only $10$ EMRI detections~\citep{GTVemri}, so the same measurement should be possible with NGO. Similarly, LISA would be able to constrain the Hubble constant to $\sim 1\%$ by the detection of $\sim 20$ EMRIs at $z<0.5$~\citep{2008PhRvD..77d3512M}, which should again be possible for NGO. In addition, any individual EMRI event that is observed can be used to place strong constraints on deviations from the no-hair property of Kerr black holes~\citep[see][for a review]{ASReview}. The scientific potential of EMRI events observed with NGO for astrophysics, cosmology and fundamental physics is therefore very strong. Further work is required to fully quantify the impact of the rescope on all of these scientific objectives.

\bibliography{GairLISA9_EMRIs}

\begin{thebibliography}{}
\expandafter\ifx\csname natexlab\endcsname\relax\def\natexlab#1{#1}\fi
\expandafter\ifx\csname url\endcsname\relax
  \def\url#1{\texttt{#1}}\fi
\expandafter\ifx\csname urlprefix\endcsname\relax\def\urlprefix{URL }\fi
\providecommand{\eprint}[2][]{\url{#2}}

\bibitem[{{Amaro-Seoane} et~al.(2007){Amaro-Seoane}, {Gair}, {Freitag},
  {Miller}, {Mandel}, {Cutler}, \& {Babak}}]{ASReview}
{Amaro-Seoane}, P., {Gair}, J.~R., {Freitag}, M., {Miller}, M.~C., {Mandel},
  I., {Cutler}, C.~J., \& {Babak}, S. 2007, Classical and Quantum Gravity, 24,
  113

\bibitem[{{Amaro-Seoane} \& {Preto}(2011)}]{2011CQGra..28i4017A}
{Amaro-Seoane}, P., \& {Preto}, M. 2011, Classical and Quantum Gravity, 28,
  094017

\bibitem[{{Amaro-Seoane} et~al.(2012)}]{2012CQGra..29l4016A}
{Amaro-Seoane}, P., et~al. 2012, Classical and Quantum Gravity, 29, 124016

\bibitem[{{Babak} et~al.(2010)}]{mldc3}
{Babak}, S., et~al. 2010, Classical and Quantum Gravity, 27, 084009

\bibitem[{Barack \& Cutler(2004)}]{Barack:2004jy}
Barack, L., \& Cutler, C. 2004, Phys. Rev. D, 69, 082005

\bibitem[{{Danzmann}(2003)}]{2003AdSpR..32.1233D}
{Danzmann}, K. 2003, Advances in Space Research, 32, 1233

\bibitem[{Finn \& Thorne(2000)}]{Finn:2000uv}
Finn, L., \& Thorne, K. 2000, Phys. Rev. D, 62, 124021

\bibitem[{{Gair}(2009)}]{gairLISA7}
{Gair}, J.~R. 2009, Classical and Quantum Gravity, 26, 094034

\bibitem[{{Gair} et~al.(2004){Gair}, {Barack}, {Creighton}, {Cutler}, {Larson},
  {Phinney}, \& {Vallisneri}}]{emrirate}
{Gair}, J.~R., {Barack}, L., {Creighton}, T., {Cutler}, C., {Larson}, S.~L.,
  {Phinney}, E.~S., \& {Vallisneri}, M. 2004, Classical and Quantum Gravity,
  21, 1595

\bibitem[{{Gair} et~al.(2010){Gair}, {Tang}, \& {Volonteri}}]{GTVemri}
{Gair}, J.~R., {Tang}, C., \& {Volonteri}, M. 2010, \prd, 81, 104014

\bibitem[{{Hopman}(2009)}]{HopmanLISA7}
{Hopman}, C. 2009, Classical and Quantum Gravity, 26, 094028

\bibitem[{{MacLeod} \& {Hogan}(2008)}]{2008PhRvD..77d3512M}
{MacLeod}, C.~L., \& {Hogan}, C.~J. 2008, \prd, 77, 043512

\bibitem[{{Ryan}(1995)}]{ryan95}
{Ryan}, F.~D. 1995, \prd, 52, 5707

\end{thebibliography}
\bibliographystyle{asp2010}

 \end{document}